\documentclass[twocolumn,nofootinbib]{revtex4}

\usepackage[T1]{fontenc}
\usepackage{amsmath}
\usepackage{esint}
\usepackage{hyperref}
\usepackage{color}
\usepackage[normalem]{ulem}

\newcommand{\ie}{i.e.,~}

\newcommand{\be}{\begin{equation}}
\newcommand{\ee}{\end{equation}}
\newcommand{\bea}{\begin{eqnarray}}
\newcommand{\eea}{\end{eqnarray}}

\renewcommand{\tilde}{\widetilde}
\renewcommand{\hat}{\widehat}

\begin{document}

 
 \title{ \boldmath Local supersymmetry and the square roots of Bondi-Metzner-Sachs supertranslations   \mbox{  }    
}
 
 \author{Oscar Fuentealba$^{1}$} 
\author{Marc Henneaux$^{1,2}$} 
\author{Sucheta Majumdar$^{1}$} 
\author{Javier Matulich$^{1}$}
\author{Turmoli Neogi$^{1}$}

\affiliation{$^{1}$Universit\'e Libre de Bruxelles and International Solvay Institutes, ULB-Campus Plaine CP231, B-1050 Brussels, Belgium}
\affiliation{$^{2}$Coll\`ege de France, 11 place Marcelin Berthelot, 75005 Paris, France}

\begin{abstract}
{\bf Abstract:}
Super-BMS$_4$ algebras -- also called BMS$_4$ superalgebras -- are graded extensions of the BMS$_4$ algebra. They can be of two different types: they can contain either a finite number or an infinite number of fermionic generators.  We show in this letter that, with suitable boundary conditions on the graviton and gravitino fields at spatial infinity, supergravity on asymptotically flat spaces possesses as superalgebra of asymptotic symmetries a (nonlinear) super-BMS$_4$ algebra containing an infinite number of fermionic generators, which we denote SBMS$_4$. These boundary conditions are not only invariant under SBMS$_4$, but also lead to a fully consistent canonical description of the supersymmetries, which have in particular well-defined Hamiltonian generators that close according to the nonlinear SBMS$_4$ algebra. One finds in particular that the graded brackets between the fermionic generators yield all the BMS$_4$ supertranslations, of which they provide therefore ``square roots''.

\end{abstract}

\maketitle
    
The study of the gravitational field at infinity has revealed the somewhat unanticipated emergence of infinite-dimensional asymptotic symmetry groups.  This phenomenon was exhibited first in the asymptotically flat context in four spacetime dimensions, where the infinite-dimensional BMS$_4$ group, which contains the Poincar\'e group of isometries of Minkowski space as a subgroup, was shown to emerge as asymptotic symmetry group at infinity \cite{Bondi:1962px,Sachs:1962wk,Sachs:1962zza,Penrose:1962ij} (for recent reviews, see  \cite{Alessio:2017lps,Ashtekar:2018lor}).  Later and independently, anti-de Sitter gravity in three  spacetime dimensions was also shown to exhibit an infinite-dimensional extension of the anti-de Sitter group \cite{Brown:1986nw}.  

While the significance of the infinite-dimensional enhancement of the anti-de Sitter algebra takes a natural place in the context of the AdS/CFT correspondence \cite{Maldacena:1997re,Aharony:1999ti}, the physical implications of the infinite-dimensional BMS algebra are still a subject of intense  study (see \cite{Ashtekar:1987tt} for earlier work, \cite{Banks:2003vp,Barnich:2009se,Barnich:2010eb} for an intriguing extension of the formalism to include super-rotations, and \cite{Strominger:2017zoo} for review and references to the more recent exciting developments that triggered the current activity).

In the quantum theory, states are naturally defined on general Cauchy hypersurfaces \cite{Tomonaga:1946zz,Schwinger:1948yk,Footnote1}.  The asymptotic symmetries are generated by operators that act on the physical Hilbert space and form a representation of the asymptotic symmetry algebra, up to possible central terms when these are algebraically permitted.  One direct access to the quantum theory is based on the Hamiltonian formalism, which closely parallels the quantum structure.  In the standard description, the classical state of the system is completely specified (including radiation if any) by the values of  the dynamical variables on Cauchy hypersufaces, which asymptote spacelike infinity. A satisfactory formulation needs a specification of the fall-off of the phase space variables at spatial infinity, which should be such that the action and the variational principle are well-defined.  The connection between symmetries and Hamiltonian generators is then given by standard theorems of classical mechanics. One finds in particular that the symmetries have a symplectic action and are captured by the moment map (possibly defined on the centrally extended algebra when central charges occur).  This close parallel with the quantum formulation is one of reasons that make the Hamiltonian formalism instructive. 

A Hamiltonian formulation of the BMS$_4$ symmetry on spacelike hypersurfaces fulfilling the above well-definedness requirement was developed in the papers \cite{Henneaux:2018cst,Henneaux:2018hdj,Henneaux:2019yax}.  This was achieved through two distinct sets of boundary conditions.    In \cite{Henneaux:2018cst}, the parity conditions  on the leading orders of the fields in an expansion at spatial infinity were taken to be different and inequivalent to the parity conditions of \cite{Regge:1974zd}, even up to a coordinate transformation.  In \cite{Henneaux:2018hdj,Henneaux:2019yax}, the parity conditions  on the leading orders of the fields  were taken to merely differ from those of \cite{Regge:1974zd} by a coordinate transformation with specified fall-off. In both cases, the BMS$_4$ group emerges as asymptotic symmetry group of the theory.  The first set of parity conditions (of \cite{Henneaux:2018cst}) represents a more drastic twist of the parity conditions of \cite{Regge:1974zd}  because diffeomorphism invariant objects, such as the Weyl tensor, obey inequivalent conditions.  

We show in this letter how to extend the Hamiltonian analysis of the asymptotic structure of gravity to cover supergravity.  It turns out that both sets of parity conditions - those of  \cite{Henneaux:2018cst} and those of \cite{Henneaux:2018hdj,Henneaux:2019yax} - admit a supersymmetric extension of the BMS$_4$ algebra with an infinite number of fermionic generators, but that those of \cite{Henneaux:2018cst}, on which we shall focus here,  lead to a superalgebra with a richer structure that those of \cite{Henneaux:2018hdj,Henneaux:2019yax}.   In particular, the graded brackets between the fermionic generators yield all BMS$_4$ supertranslations and not just the ordinary spacetime translations.   The fermionic generators may be considered for that reason as being the ``square roots'' of the BMS$_4$ supertranslation generators. 

Earlier work on the supersymmetric extensions of the BMS$_4$ algebra considered fermionic extensions with only a finite number of fermionic generators - the standard supercharges - both at null infinity \cite{Awada:1985by} and spatial infinity  \cite{Henneaux:2020ekh}.   Extensions involving an infinite number of fermionic generators have been studied recently at null infinity \cite{Avery:2015iix,Fotopoulos:2020bqj,Narayanan:2020amh}, mostly in the context of ``celestial CFT''.  Our work differs from those interesting investigations in that we deal with spatial infinity and insist throughout in having a well-defined moment map.  We denote by SBMS$_4$ the supersymmetric extension  of the BMS$_4$ algebra with an infinite number of fermionic generators that emerges in our work.

A noticeable feature of  SBMS$_4$ is that it arises as a nonlinear superalgebra,  with  the natural definition of the supersymmetry transformations outlined below.  Nonlinearities also appear in the discussion of the asymptotic symmetries of AdS supergravity in three dimensions, where the nonlinear superalgebras of  \cite{Knizhnik:1986wc,Bershadsky:1986ms,Fradkin:1992bz} emerge at infinity \cite{Henneaux:1999ib}, but we have not analyzed in the SBMS$_4$ case whether the nonlinearities were intrinsic or could be redefined away. 

The Hamiltonian action of simple supergravity in four spacetime dimensions reads \cite{Teitelboim:1977fs,Tabensky:1977ic,Fradkin:1977wv,Deser:1977ur,Pilati:1977ht}
 \be
S[\pi_{a}^{i},e_{i}^{a},\psi_{m},\omega; N, N^k, \psi_0, \lambda_{ab}] = 
 \int dt\left[K-H\right] \,, \label{eq:BulkAction0}
\ee
where the kinetic term is
\begin{equation}
K=\int d^{3}x\left(\pi_{a}^{i}\dot{e}_{i}^{a}+\frac{i}{2}\sqrt{g}\psi_{k}^{T}\gamma^{km}\dot{\psi}{}_{m}\right) + B_\infty^K\, \label{eq:Kinetic00}
\end{equation}
and where the Hamiltonian is
\be
H = \int d^{3}x\left(N\mathcal{H}+N^{i}\mathcal{H}_{i}+i\psi_{0}^{T}\mathcal{S}+\frac{1}{2}\lambda_{ab}\mathcal{J}^{ab}\right) + B_\infty^H\,. 
\ee
Here, $B_\infty^K$ and $B_\infty^H$ are surface integrals at infinity.  The explicit form of $B_\infty^K$ will be determined below, while $B_\infty^H$ is the standard ADM energy ($N \rightarrow 1$, $N^k \rightarrow 0$ at infinity). The canonical variables are the triads $e^a_i$, their conjugate momenta $\pi^i_a$ and the components $\psi_k$ of the gravitino field.  They also include the field $\omega$, which is a fermionic surface field that must be introduced at infinity to preserve invariance under Lorentz boosts subject to the relaxed boundary conditions that we shall adopt following \cite{Fuentealba:2020aax}.  This field enters only in the surface term $B^K_\infty$. 

The lapse $N$ and the shift $N^k$ are the Lagrange multipliers associated to the Hamiltonian and momentum constraints,
whose explicit expressions are 
\begin{eqnarray}
\mathcal{H} & = & \frac{1}{\sqrt{g}}\left(\pi_{ij}\pi^{ij}-\frac{1}{2}\pi^{2}\right)-\sqrt{g}R+F_{1} \approx 0\,,\\
\mathcal{H}_{i} & = & -\nabla_{j}\pi_{i}^{j}+F_{2} \approx 0\,,
\end{eqnarray}
with $g_{ij} = e^a_i e_{aj}$, $\pi^{ij}=e^{a(i}\pi_{a}^{j)}$ \cite{Footnote-factor2}. Here, $F_{1}$ and
$F_{2}$ are bilinear in fermions ($\sim\psi\partial\psi,\,\omega\psi\psi,\,\pi\psi\psi$, where $\omega$ stands for the spatial spin connection $\omega_{abk}$ and $\pi$ stands for the momenta $\pi^i_a$).
The time component of the gravitino field $\psi_{0}$ plays the role
of the Lagrange multiplier associated to the fermionic constraint
\begin{equation}
\mathcal{S}=\sqrt{g}\gamma^{mn}\partial_{m}\psi_{n}+F_{3}+F_{4} \approx 0 \,,\label{eq:ContraintS}
\end{equation}
where $F_{3}$ is linear in fermions  with coefficients that involve the spatial spin connection or the triad momenta ($\sim\omega\psi,\,\pi\psi$) and
$F_{4}$ is at least cubic in fermions. The constraints $\mathcal{J}_{ab} =-  \mathcal{J}_{ba}$  read 
\be
\mathcal{J}^{ab} = \pi^{ai}e^b_i -  \pi^{bi}e^a_i+ \frac12 \sqrt{g} \psi^T_i \gamma^{ij} \gamma^{ab} \psi_j\approx 0
\ee
and do not involve derivatives of the fields.  They generate local spatial rotations and the $\lambda_{ab}$'s are their respective
Lagrange multipliers.

We freeze asymptotically the freedom of making rotations of the local frames by tying at infinity these local frames  to the cartesian coordinates, i.e. $\{e^a \equiv e^a_i dx^i \}\rightarrow \{dx^a\}$ or equivalently for the dual basis, $\{e_a \}\rightarrow \{\frac{\partial}{\partial x^a}\}$, more precisely,
\begin{eqnarray}
e_{i}^{a} & = & \delta_{i}^{a}+\frac{1}{2}\delta^{aj}h_{ij}+\mathcal{O}(r^{-2})\,,\label{eq:FO-e}\\
\pi_{a}^{i} & = & \delta_{aj}\pi^{ij}+\mathcal{O}(r^{-3})  \, ,
\end{eqnarray}
where $h_{ij}$ and $\pi^{ij}$ behave as in \cite{Henneaux:2018cst}, namely,
\begin{eqnarray}
h_{ij} & = & \frac{\overline{h}_{ij}(\theta, \varphi)}{r}+\mathcal{O}(r^{-2})\,,\\
\pi^{ij} & = & \frac{\overline{\pi}^{ij}(\theta, \varphi)}{r^{2}}+\mathcal{O}(r^{-3})\,,
\end{eqnarray}
with leading orders subject to definite parity conditions explicitly spelled out in that article \cite{Henneaux:2018cst,Compere:2011ve}. These boundary conditions on the bosonic fields imply in particular that $\pi^i_a$ and the spatial spin connection $\omega_{abk}$ decrease at least as $r^{-2}$ at infinity.  

The crucial steps that lead to the extension of the BMS$_4$ algebra with an infinite number of fermionic generators, all acting as well-defined canonical transformations, are as follows.

\textbf{(1)} First, one extends the boundary conditions of  \cite{Henneaux:2020ekh} on the fermionic fields along the lines of \cite{Fuentealba:2020aax}, by adding to the standard $\mathcal{O}(r^{-2})$ behaviour a gradient term that decays slowlier at infinity, $\psi_{k}  =  \partial_k \chi + \mu'_k$ with $ \chi = \mathcal{O}\left(1\right)$ and $\mu'_{k} =\mathcal{O}(r^{-2})$.  The term $\partial_k \chi$ is the leading piece of a supersymmetry transformation of the gravitino field $\delta_\chi \psi_k$.   The subsequent terms in  $\delta_\chi \psi_k$ are of lower order and can be included in $\mu'_k$. The same is true for the supersymmetry variations of the bosonic fields, which are of the same type as the existing terms $\overline{h}_{ij}(\theta, \varphi)/r$ and  $\overline{\pi}^{ij}(\theta, \varphi)/r^{2}$, as it will indirectly follow from our subsequent analysis. The idea of the new extended boundary conditions, thus, is that the supergravity fields are defined at spatial infinity up to a supersymmetry transformation.  Because this supersymmetry transformation has a non-vanishing charge, it is an improper gauge transformation with non trivial physical content, and it cannot be gauged away \cite{Benguria:1976in}.  Once included, it must be kept.

In order to maintain manifest space covariance, we write, instead of  the decomposition $\psi_{k}  =  \partial_k \chi + \mu'_k$, the equivalent  covariant decomposition
\be
\psi_{k}  =  \nabla_k \chi + \mu_k, \quad \chi = \mathcal{O}\left(1\right), \quad \mu_{k} =\mathcal{O}\left(r^{-2}\right)\,, \label{Eq:DecayPsi}
\ee
the difference between $\nabla_k \chi$ and $\partial_k \chi$ being absorded in a redefinition of the subleading term $\mu_k$. 
The spinor $\chi$ is moreover assumed to be even to leading order, \be \chi(x^i) = \chi^{(0)}(n^i) + \frac{\chi^{(1)}(n^i)}{r} + \mathcal{O}(r^{-2}), 
\ee
with
\be
 \chi^{(0)}(-n^i) = \chi^{(0)}(n^i).
\ee
Here, $n^i$ stands for the unit normal to the sphere at infinity, or what is the same, the angles $(\theta, \varphi)$ on the $2$-sphere at infinity.  Furthermore, since the leading order of $\psi_{k}$ is given by $\partial_k \chi^{(0)}$, we can assume that  $ \chi^{(0)}$ has no zero mode \cite{Extra}.  

\textbf{(2)} With these boundary conditions, the kinetic term in the action is formally divergent.  However, the leading (linear) divergence is actually a total derivative, and so can be eliminated by adapting the surface term $B^K_\infty$ as $B^K_\infty = \tilde{B}_\infty^{div} + \tilde{B}_\infty^K$ where $\tilde{B}_\infty^{div} $ cancels the bulk linear divergence  and $\tilde{B}_\infty^K$ will be given below.  The next apparent divergence is logarithmic but is actually not present with the parity conditions given above.  

To display the manifestly finite form of the kinetic term, it is convenient to extend $\chi$ and $\mu_k$ in the bulk and treat them as independent bulk fields in the variational principle.  This is permissible since the equations obtained by varying $\chi$ and $\mu_k$ are equivalent to the original equations obtained by varying $\psi_k$.  The procedure merely introduces an extra redundancy in the description, given by $\chi \rightarrow \chi + \sigma$, $\mu_k \rightarrow \mu_k - \nabla_k \sigma$, under which the field $\psi_k$ is invariant.  To remove the degeneracy of the symplectic term, one introduces at the same time the momentum $\pi_\chi$ conjugate to $\chi$ in the bulk and the fermionic first class constraint 
\be
\mathcal{F}  = \pi_\chi +  \mathcal{M} \approx 0\,,
\ee
which canonically generates this redundancy,  together with its Lagrange multiplier $\Lambda$.  Here, $\mathcal{M}$ stands for the coefficient of $\dot{\chi}$ in the expression obtained by substituting $\psi_{k}  =  \nabla_k \chi + \mu_k$ in $\frac{i}{2}\sqrt{g}\psi_{k}^{T}\gamma^{km}\dot{\psi}$ and making the integration by parts necessary to get rid of the time derivative $\dot{\mu}_k$. Again this step is permissible since $\pi_\chi$ and $\Lambda$ can be viewed as auxiliary fields that can be eliminated using their own equations of motion, yielding back the formulation without $\pi_\chi$ and $\Lambda$.

When all this is done, the kinetic term takes the explicit form
\begin{equation}
K=\int d^{3}x\left(\Pi_{a}^{i}\dot{e}_{i}^{a}+ i \pi_\chi^T \dot{\chi} + \frac{i}{2}\sqrt{g}\mu_{k}^{T}\gamma^{km}\dot{\mu}{}_{m}\right) + \tilde{B}_\infty^K\, \label{eq:Kinetic01}
\end{equation}
with 
\be
\tilde{B}_\infty^K = \frac{i}{2} \oint d \theta d \varphi \sin \theta \omega^T  \dot{\chi}^{(0)}.
\ee
As we recalled above, the surface field $\omega$,  which is conjugate to the asymptotic value of $\chi$, is necessary for the canonical implementation of Lorentz invariance.  This was shown in \cite{Fuentealba:2020aax} for the linear theory but the argument can be straightforwardly be extended to the interacting one.  Without loss of generality, we take $\omega$ to be even and without zero mode. 

The fields $\Pi^i_a$ involve a parity-preserving redefinition of the variables $\pi^i_a$ that make them have simpler canonical brackets with the other phase space variables. This redefinition follows from the above decomposition of $\psi_k$ and the subsequent redefinitions in the kinetic term.  Note, however, that the $\Pi^i_a$'s  are not invariant under the redundancy $\chi \rightarrow \chi + \sigma$, $\mu_k \rightarrow \mu_k - \nabla_k \sigma$, while the original $\pi^i_a$ are.  

The action takes the form
\be
S[\Pi_{a}^{i},e_{i}^{a},\psi_{m},\omega; N, N^k, \psi_0, \lambda_{ab}] = 
 \int dt\left[K-H\right] \,, \label{eq:BulkAction01}
\ee
where $K$ is given by (\ref{eq:Kinetic01}) and $H$ involves the new constraint,
\be
H = \int d^{3}x\Big(N\mathcal{H}+N^{i}\mathcal{H}_{i}+i\psi_{0}^{T}\mathcal{S}+\frac{1}{2}\lambda_{ab}\mathcal{J}^{ab}+ i\Lambda \mathcal{F} \Big) + B_\infty^H\,.
\ee

\textbf{(3)} Poincar\'e invariance follows straightforwardly from the analysis  of \cite{Henneaux:2018cst}  and \cite{Fuentealba:2020aax}.  This is because the interaction terms are subleading at spatial infinity with respect to the free terms.  We focus therefore on supersymmetry. 

The functional
\be
G_\epsilon = \int d^3x (\epsilon^T \mathcal{S} +  \xi^i(\epsilon) \mathcal{H}_i + i \lambda^T(\epsilon) \mathcal{F} )+ \Sigma_\epsilon\,,
\ee
where the supersymmetry parameter
\be
\epsilon = \mathcal{O}(1) = \epsilon^{(0)}(n^i) + \mathcal{O}(1/r)
\ee
is such that its leading order is an arbitary even function $\epsilon^{(0)}(-n^i) = \epsilon^{(0)}(n^i) $ on the 2-sphere, is a well defined canonical generator, in the sense that it fulfills $\iota_X \Omega = - d_V G_\epsilon$ for some vector field $X$ (exactly, and not up to surface terms).   Here, $\Omega$ is the symplectic form, $d_V$ the exterior derivative in field space and $\iota_X \Omega$ the inner contraction of $\Omega$ by the vector field $X$ defining the transformation generated by $G_\epsilon$.   

The $\epsilon$-dependent compensating diffeomorphisms and reshufflings between $\chi$ and $\mu$ generated by $ \xi^i(\epsilon) \mathcal{H}_i + \lambda^T(\epsilon) \mathcal{F}$ are included to preserve the boundary conditions (specifically $\bar{h}_{rA} = 0$ and $\mu_i = \mathcal{O}(r^{-2})$, where $\bar{h}_{rA}$ is the  radial-angular component of the leading order of the metric \cite{Henneaux:2018cst}, as well as the absence of zero-mode condition for $\chi^{(0)}$ and $\omega$). One has (to leading $\mathcal{O}(1)$ order)
\be
 \xi^i (\epsilon) = -\frac{i}{2}\partial^{i}\left(r\epsilon^{T}\gamma_{0}\gamma_{r}\right)\chi\,, \quad \lambda (\epsilon) = -\big(\epsilon - \epsilon_{0}\big)\,,
\ee
where $\epsilon_0$ stands for the zero mode of the parameter $\epsilon$. Finally,  the surface term $\Sigma_\epsilon$ to which the generator $G_\epsilon$ reduces when the constraints hold is given by
\begin{align}
\Sigma_{\epsilon} & =-i\oint d^{2}x\sqrt{\overline{g}}\epsilon_{0}^{T}\gamma_{r}\overline{\gamma}^{A}\overline{\mu}_{A} \nonumber\\
&+\frac{i}{4}\oint d^{2}x\sqrt{\overline{g}}\left(\epsilon+\epsilon_{0}\right)^{T}\chi^{(0)}\overline{h}_{rr} \nonumber \\
 & \quad-\frac{i}{8}\oint d^{2}x\sqrt{\overline{g}}\left(\epsilon-\epsilon_{0}\right)^{T}\gamma_{r}\overline{\gamma}^{A}\chi^{(0)}\partial_{A}\overline{h}_{rr} \nonumber \\
 & -\frac{i}{2}\oint d^{2}x\sqrt{\overline{g}}\left(\epsilon-\epsilon_{0}\right)^{T}\omega \nonumber \\
 & \quad-\frac{i}{2}\oint d^{2}x\epsilon^{T}\gamma_{0}\gamma_{r}\chi^{(0)}\left(\overline{\Pi}^{rr}-\overline{\Pi}_{A}^{A}\right)\,,  \label{eq:SigmaEpsilon}
\end{align}
where $d^2x=d\theta d\varphi$ corresponds to the integral measure and $\overline{g}$ stands for the determinant of the metric on the $2$-sphere at infinity.  The surface integral $\Sigma_{\epsilon}$ involves both linear and quadratic terms in the asymptotic fields. Note that in this integral only the asymptotic value of $\epsilon$, \ie $\epsilon^{(0)}$ contributes.

The transformation generated by $G_\epsilon$ is a standard supersymmetry transformation with supersymmetry parameter $\epsilon$. Indeed, one can see for instance that the transformation of the metric and the gravitino field reads
 \be
 \delta_\epsilon g_{ij} = i \bar{\epsilon} \gamma_{(i}\psi_{j)} + ``more" , \quad \delta_\epsilon \psi_i = -\!^{(4)}\nabla_i \epsilon + ``more"
 \ee
where $``more"$ in the metric transformation law stands for the coordinate transformation that is included to preserve $\overline{h}_{rA} = 0$, while $``more"$ in the gravitino transformation law stands for higher order terms in fermions. This implies in particular
\be
\delta_{\epsilon} \chi^{(0)} = -\big(\epsilon - \epsilon_{0}\big)\,.
\ee
The conjugate to $ \chi^{(0)}$ transforms as
\be
\delta_{\epsilon} \omega =\frac{1}{2}\big(\epsilon + \epsilon
_{0}\big)\overline{h}_{rr} +\frac{1}{4}\gamma_r \overline{\gamma}^A\big(\epsilon - \epsilon_{0}\big)\partial_A\overline{h}_{rr}\,,
\ee
insuring that the transformation is canonical (no unwanted surface term in the relation $\iota_X \Omega = - d_V G_\epsilon$ relating the exterior derivative in field space of the generator $G_\epsilon$ to the contraction of the symplectic form $\Omega$ with the vector field $X$ defining the infinitesimal transformations).

The transformations with non-vanishing $\epsilon$ at infinity are improper gauge symmetries since the surface term (\ref{eq:SigmaEpsilon}) does not vanish in this case \cite{Benguria:1976in}.   Because the asymptotic value of the supersymmetry parameter is an arbitrary (even) function $\epsilon^{(0)}(\theta, \varphi)$  on the $2$-sphere, we have given a formulation of supergravity with an infinite number of improper fermionic gauge symmetries. 

The set of fermionic transformations is in fact larger since the theory is also invariant under time-independent shifts of $\omega$ with generator $Q_\sigma$
\be
\delta_\sigma \omega = \sigma = [\omega , Q_\sigma]\,, \quad Q_\sigma  = \frac{i}{2} \oint d^2 x \sqrt{\overline{g}} \,  \sigma^T \, \chi^{(0)}\,. 
\ee
(Note that the equations of motion imply in particular $\dot{\chi}^{(0)} = 0$ and hence $\dot{Q}_\sigma= 0$.)

\textbf{(4)} The asymptotic symmetry algebra is the standard one for Poincar\'e and BMS, in the Hamiltonian basis \cite{Troessaert:2017jcm,Henneaux:2018cst}. The two additional fermionic symmetries $G_\epsilon$ and $Q_\sigma$ form infinite-dimensional (reducible but indecomposable) representations of the homogeneous Lorentz group, with zero modes transforming in the finite-dimensional spin-$\frac12$ representation.  In the case of $G_\epsilon$, there are additional nonlinear terms in the bracket with the boost generators, involving $Q_\sigma$ and the supertranslations, due to the quadratic terms in the surface integral $\Sigma_\epsilon$.
The fermionic symmetries both commute with the supertranslations.

\textit{Finally, the (graded) brackets of the fermionic symmetries provide square roots of all BMS supertranslations, \ie $[G_\epsilon, G_{\epsilon'}]\sim G_{\hat{T},\hat{W}}$, generalising the familiar relation $[Q,Q'] \sim \gamma^\mu P_\mu$ of ordinary finite-dimensional supersymmetry, where the resulting BMS supertranslation parameter is given by}
\begin{align}
\hat{T}& =-\frac{i}{4}\epsilon^T \epsilon' \,,\\
\hat{W} &= \frac{i}{4} \epsilon^T \gamma_0 \gamma_r \epsilon'-\frac{i}{4}\left(\epsilon^T \gamma_0\gamma_r\epsilon'_
{0}-\epsilon'^T \gamma_0\gamma_r\epsilon_
{0}\right) \,.
\end{align}
$T$ and $W$ are even and odd functions under parity, respectively, and together form the angle-dependent supertranslation parameter \cite{Henneaux:2018cst,Henneaux:2019yax}.
The brackets of the other fermionic symmetries vanish, $[Q_\sigma, Q_{\sigma'}]= 0$, while there is a non-trivial central charge in the bracket $[G_\epsilon, Q_\sigma]=-\frac{i}{2}\oint d^{2}x\sqrt{\overline{g}}\left(\epsilon-\epsilon_{0}\right)^{T}\sigma$.

\vspace{.2cm}

We have thus successfully provided boundary conditions at spatial infinity for supergravity that are invariant under an extension of the BMS$_4$ algebra by an infinite number of fermionic generators.  These can be thought of as square roots of the supertranslations.    One central idea in the construction is to enlarge the boundary conditions of \cite{Henneaux:2020ekh}, which lead to a finite-dimensional fermionic extension of the BMS$_4$ algebra, by allowing an improper gauge term $\partial_k \chi$ that decays slowlier at infinity in the gravitino field $\psi_k$.

Extending the formalism in order to enlarge the set of improper gauge transformations may not be always direct or possible, as the papers \cite{Tanzi:2020fmt,Tanzi:2021xva} have shown for the minimal electromagnetic coupling.  What makes the extension work in the case of supergravity is the fact that the abelian part of the supersymmetry transformation dominates the non abelian corrections, as for pure gravity \cite{Fuentealba:2020ghw}, while this is not so for Yang-Mills theory where the derivative operator $\partial_k$ and the connection $A_k$ are of same order $\mathcal{O}(r^{-1})$.   The difficulty is somewhat reminiscent of the ``boost problem'' in field theories \cite{Christodoulou1981} and also of the difficulties encountered in imposing asymptotically the Lorenz gauge in the case of electrodynamics with massless charged fields \cite{Tanzi:2021xva,Satishchandran:2019pyc}.  

One attractive feature of the Hamiltonian description of the symmetries on Cauchy hypersurfaces is that it does not rely on the dynamical question of the existence of a smooth null infinity, which is a delicate and somewhat intricate issue \cite{Damour:1985cm,Christodoulou1,Friedrich2004,ValienteKroon:2002fa,Friedrich:2017cjg,Kehrberger:2021uvf,Kehrberger:2021vhp}.  

Two intriguing properties of the superalgebra SBMS$_4$ should be stressed.  First, nonlinear terms appear in the brackets of the asymptotic generators, as in AdS$_3$ supergravity \cite{Henneaux:1999ib}, but we have not explored whether these nonlinear terms could be absorbed through redefinitions.  Second,  another infinite-dimensional fermionic symmetry arises, generated by $Q_\sigma$. It would be worthwhile to understand the reason behind its emergence, which might perhaps be related to sub-leading soft theorems through the corresponding Ward identities (see \cite{Strominger:2017zoo}  for the connection between soft theorems and asymptotic symmetries).

Since we derived a bone fide Hamiltonian description of the asymptotic symmetry, the transition to the quantum theory can be performed by applying the usual correspondence rules.  The symmetry generators would correspond to quantum operators, with an algebra that is $(i \hbar)$ times the classical bracket algebra up to possible corrections of higher order in Planck's constant.   The fact that the supertranslations are given by the anticommutators of fermionic symmetries might lead to interesting new positivity theorems \cite{Teitelboim:1977hc,Deser:1977hu} (provided the Hilbert space has no negative norm states \cite{Boulware:1985nn}).  

The detailed derivation of the results presented here, as well as the discussion of other boundary conditions, will be given in a separate publication.

We thank Ricardo Troncoso for useful discussions. This work was partially supported by the ERC Advanced Grant “High-Spin-Grav”, by FNRS-Belgium (conventions FRFC PDRT.1025.14and IISN 4.4503.15), as well as by funds from the Solvay Family.


\begin{thebibliography}{99}
\bibitem{Footnote1}  If the spacetime fails to be globally hyperbolic, there is no Cauchy hypersurface.  This is not the case for Minkowski space (with or without radiation fields) but the phenomenon might occur, however, for more general spacetimes.  Our considerations still apply then in a neighbourhood of spatial infinity - more precisely in a causal diamond containing spatial infinity and appropriate portions of past and future null infinity ``touching'' it.
\bibitem{Extra}
On the technical side, we complete the boundary conditions by also imposing, as in the bosonic case, that the constraint functions decay two orders faster than the one implied by the behaviour of the fields, i.e.,
$\mathcal{H} =  \mathcal{O}(r^{-5})$,  $\mathcal{H}_i =  \mathcal{O}(r^{-5})$ and $\mathcal{S} =  \mathcal{O}(r^{-5})$  (in Cartesian coordinates).  For the Lagrange multipliers, we assume that the time evolution is a time translation at infinity which involves no shift or supersymmetry transformation, i.e., $N =  1 + \mathcal{O}(r^{-1})$, $N^i =  \mathcal{O}(r^{-1})$ and $ \psi_0 =  \mathcal{O}(r^{-1})$.
\end{thebibliography}
\end{document}